\documentclass[times,twocolumn,final]{elsarticle}

\usepackage{medima}
\usepackage{framed,multirow}

\usepackage{amssymb}
\usepackage{latexsym}

\usepackage{url}
\usepackage{xcolor}

\usepackage{hyperref}

\usepackage{amsmath,amssymb,amsfonts}
\usepackage{algorithmic}
\usepackage{graphicx}
\usepackage{textcomp}
\usepackage{float}
\usepackage{booktabs}
\usepackage{natbib}
\usepackage{color}
\usepackage{tikz}
\usetikzlibrary{shapes}

\definecolor{newcolor}{rgb}{.8,.349,.1}

\journal{Medical Image Analysis}

\newcommand{\myRed}[1]{\textcolor{black}{#1}}

\begin{document}

\verso{Sergey Kastryulin \textit{et~al.}}

\begin{frontmatter}

\title{Image Quality Assessment for Magnetic Resonance Imaging}%

\author[1,2]{Sergey \snm{Kastryulin}}
\author[2]{Jamil \snm{Zakirov}}
\author[3,4]{Nicola \snm{Pezzotti}}
\author[2]{Dmitry V. \snm{Dylov}\corref{cor1}}
\ead{d.dylov@skoltech.ru}
\cortext[cor1]{Corresponding author:}

\address[1]{Philips Research, Moscow, Russia}
\address[2]{Skolkovo Institute of Science and Technology, Moscow, Russia}
\address[3]{Philips Research, Eindhoven, The Netherlands}
\address[4]{Eindhoven University of Technology, Eindhoven, The Netherlands}


\begin{abstract}
Image quality assessment (IQA) algorithms aim to reproduce the human's perception of the image quality.  
The growing popularity of image enhancement, generation, and recovery models instigated the development of many methods to assess their performance. 
However, most IQA solutions are designed to predict image quality in the general domain, with the applicability to specific areas, such as medical imaging, remaining questionable.
Moreover, the selection of these IQA metrics for a specific task typically involves intentionally induced distortions, such as manually added noise or artificial blurring; yet, the chosen metrics are then used to judge the output of real-life computer vision models. 
In this work, we aspire to fill these gaps by carrying out the most extensive IQA evaluation study for Magnetic Resonance Imaging (MRI) to date (14,700 subjective scores).
We use outputs of neural network models trained to solve problems relevant to MRI, including image reconstruction in the scan acceleration, motion correction, and denoising.
Our emphasis is on reflecting the radiologist's perception of the reconstructed images, gauging the most diagnostically influential criteria for the quality of MRI scans: signal-to-noise ratio, contrast-to-noise ratio, and the presence of artifacts.
Seven trained radiologists assess these distorted images, with their verdicts then correlated with 35 different image quality metrics (full-reference, no-reference, and distribution-based metrics considered).
\myRed{The top performers -- DISTS, HaarPSI, VSI, and FID\textsubscript{VGG16} -- are found to be efficient across three proposed quality criteria, for all considered anatomies and the target tasks.}
\end{abstract}

\begin{keyword}
\MSC 68T45\sep 62P10
\KWD Image~Quality\sep Deep~Learning\sep Metrics\sep Reconstruction~Quality\sep Magnetic~Resonance~Imaging
\end{keyword}

\end{frontmatter}


\section{Introduction}
\label{sec:introduction}

Image quality assessment (IQA) is a research area occupied with constructing accurate computational models to predict the perception of image quality by human subjects, the ultimate consumers of most image processing applications \cite{wang2011applications}.

The growing popularity of image enhancement and image generation algorithms increases the need for a quality assessment of their performance. The demand has led to the abundance of IQA methods emerging over the last decades. The well-known \textit{full-reference} (FR) metrics, such as MSE, PSNR, and SSIM \cite{wang2004image, wang2006modern}, became a de-facto standard in many computer vision applications.
The more recent \textit{no-reference} (NR) metrics, such as BRISQUE \cite{MittalMB12}, have also found their use, especially when the ground truth images are absent or hard to access.
Yet, another class of \textit{distribution-based} metrics (DB) earned the community's attention, thanks to the advent of generative adversarial networks (GANs), enabling the quality assessment using distributions of thousands of images instead of gauging them individually.
The popular new DB IQA methods include such metrics as Inception Score \cite{salimans2016improved}, FID \cite{heusel2017gans}, KID \cite{binkowski2018demystifying}, MSID \cite{tsitsulin2020shape}, and many others.
%
Despite being widely used, the DB metrics were neither included in the recent large scale general domain reviews \cite{athar2019comprehensive, manap2015non}, nor in the medical ones\cite{Mason2020medicalcomparison}.

IQA measures are applied to estimate the quality of image processing algorithms and systems.
For example, when several image denoising and restoration algorithms are available to recover images distorted by blur and noise contamination, a perceptual objective IQA could help pick the one that generates the best perceptual image quality after the restoration.
To do that reliably, Image Quality Metrics (IQMs) need to show a high correlation with the perceptual estimates of the quality reported by human subjects for a given image processing algorithm. 
However, IQA algorithms are often \textit{evaluated on non-realistic distortions}, such as added noise or artificial blurring\cite{athar2019comprehensive, manap2015non,Mason2020medicalcomparison}.
Such a discrepancy between the synthetic evaluation and the practical use may cause misleading results.

While most metrics are designed to predict image quality in the general domain,  Magnetic Resonance Imaging (MRI) provides gray-scale data with the content and style noticeably different from the natural images.
Hence, the applicability of the IQMs in the MRI domain must be validated.

Moreover, IQMs trained on natural images attempt to describe the overall perception of the quality of an entire scene. 
On the contrary, an MRI scan can be perceived as high-quality when specific characteristics, responsible for the scan's value, are deemed adequate. 
Those are the characteristics that assist expert radiologists in making their diagnostic decisions, including perceived level of noise (signal-to-noise ratio, SNR), perceived soft tissue contrast (contrast-to-noise ratio, CNR), and the presence of artifacts.
These specific quality criteria are, unfortunately, coupled.
For example, some denoising algorithms tend to introduce additional blurring (lowering of CNR) in exchange for increased SNR, and some motion correction approaches tend to introduce noticeable artifacts.
Therefore, a more detailed evaluation of IQM's ability to express separate MRI quality criteria is required.

\section{Related Work}
\label{sec:related-work}

The evaluation of metrics for IQA in the domain of natural images started from the early task-specific works that considered FR methods to characterize color displays and half-toning optimization methods \cite{Ahumada1993}.

More recent task-specific studies explored IQA for the images of scanned documents \cite{Kumar2013} and screen content \cite{Yang2015}.
Likewise, fused images \cite{Wang2021}, smartphone photographs~\cite{Fang2020}, remote sensing data \cite{Ieremeiev2020}, and climate patterns \cite{Baker2019} demanded the development of targeted IQA approaches.
Historically, many of these works have been focusing on the quality degradation caused by the compression algorithms \cite{avadhanam1999evaluation, allen2007image, triantaphillidou2007image, Baker2019}, with relatively small datasets appearing publicly for the IQ evaluation.
However, the small dataset size and the excessive re-use of the same test sets have led to the promotion of the IQMs poorly generalizable to the unseen distortions.

This was recognized as a major problem, stimulating the emergence of large-scale studies \cite{athar2019comprehensive, ding2020comparison}.
Among the large-scale evaluations, the majority compared multiple FR metrics, ranging from just a handful \cite{Sheikh2006, Larson2010, Zhang2012, Mohammadi2014} to several dozens \cite{athar2019comprehensive,Pedersen2015comprehensive} of IQMs analyzed on popular datasets.

\medskip
The medical domain stands out from the others by a special sense of what is deemed informative and acceptable in the images \cite{Cavaro-Menard2010challenges}. 
Resulting from years of training and practice, the perception of medical scan quality by adept radiologists relies on a meticulous list of anatomy-specific requirements, on their familiarity with particular imaging hardware, and even on their intuition.

Given the majority of IQMs were not designed for the healthcare domain, some recent works were dedicated to the niche. 
One small-scale study considered a connection of IQA of natural and medical images via signal-to-noise (SNR) estimation \cite{Li2018naturaltomedical}.
%
Others assessed common FR IQMs using non-expert raters \cite{Rajagopal2015nonexpert, Chow2016nonexpert}.
Sufficient for the general audience, these methods proved incapable of reflecting the fine-tuned perception of the radiologists \cite{Keshavan2019expertassessment}.

Expert raters were then engaged in \cite{Renieblas2017medicalssim} and \cite{Mason2020medicalcomparison}. 
The former studied only IQMs from the SSIM family and the latter assessed 10 FR IQMs, reporting that VIF \cite{sheikh2005visual}, FSIM \cite{zhang2011fsim}, and NQM \cite{Damera-Venkata2000NQM} yield the highest correlation with the radiologists' opinions.

On the other hand, Crow \textit{et al.} argue that NR IQA are preferable for assessing medical images because there may be no perfect reference image in the real-world medical imaging \cite{Chow2016biomedialIQAreview}.
To address this issue, several recent studies also propose new NR IQMs for MRI image quality assessment \cite{Jang2018nriqa, Oszust2020nriqm, Obuchowicz2020nriqm, Stepien2021nriqm, Simi2021nriqm, kustner2018mrnriqm, mortamet2009automatic, pizarro2016automated}.
Unfortunately, none of these new metrics have an open-source implementation, making verification of the claimed results problematic.

\section{Image Quality Metrics Considered}
In this work, we evaluate the most widely used and publicly available general-purpose FR, NR, and DB IQMs to find the best algorithms for the quality assessment on arguably the most important MRI-related image-to-image tasks: \textit{scan acceleration}, \textit{motion correction}, and \textit{denoising}. 
Instead of modeling the disrupted images, we use outputs of trained neural networks and compare them with the clean reference images from the fastMRI dataset.

Our study includes the following 35 metrics: \textbf{17 Full-Reference IQMs} (PSNR, SSIM \cite{wang2004image}, MS-SSIM \cite{wang2003multiscale}, IW-SSIM \cite{wang2010information}, VIF \cite{sheikh2005visual}, GMSD \cite{xue2013gradient},  MS-GMSD \cite{Zhang2017gmsd}, FSIM \cite{zhang2011fsim}, VSI \cite{zhang2014vsi}, MDSI \cite{nafchi2016mdsi}, HaarPSI \cite{Reisenhofer2018haarpsi}, Content and Style Perceptual Scores \cite{johnson2016perceptual}, LPIPS \cite{zhang2018unreasonable}, DISTS \cite{ding2020image}, PieAPP \cite{Prashnani2018pieapp}, DSS \cite{Balanov2015dss}), 
\textbf{3 No-Reference IQMs }(BRISQUE \cite{MittalMB12}, PaQ-2-PiQ \cite{paq2piq2020}, MetaIQA \cite{MetaIQA2020}),
and \textbf{15 Distribution-Based IQMs} (KID \cite{binkowski2018demystifying}, FID \cite{heusel2017gans}, GS \cite{khrulkov2018geometry}, Inception Score (IS) \cite{salimans2016improved}, MSID \cite{tsitsulin2020shape}, all implemented with three different feature extractors: Inception Net \cite{szegedy2015going}, VGG16, and VGG19 \cite{simonyan2014very}).
For brevity of the presentation, we will showcase only the analysis of the best performing four metrics, in the order of their ranking: VSI \cite{zhang2014vsi}, HaarPSI \cite{Reisenhofer2018haarpsi}, DISTS \cite{ding2020image}, and FID\textsubscript{VGG16} \cite{heusel2017gans}.
\myRed{
All metrics were re-implemented in Python to enable a fair comparison, with the PyTorch Image Quality (PIQ) \cite{piq} chosen as the base library for implementing all metrics.
The resulting implementations were verified to be consistent with the original implementations proposed by the authors of each metric.
}

For the comparison, we collect 14,700 ratings from 7 trained radiologists to evaluate the quality of reconstructed images based on three main criteria of quality: perceived level of noise (SNR), perceived soft tissue contrast (CNR), and the presence of artifacts, making this work the most comprehensive study of MRI image quality assessment to date.
\section{Medical Evaluation}
\label{sec:medical-evaluation}
\begin{figure*}[!ht]
    \centering
    \includegraphics[width=\linewidth]{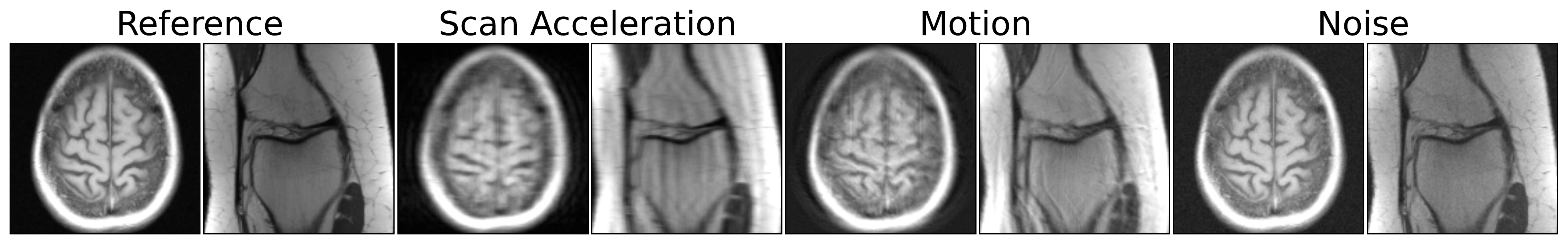}
    \caption{Distortions introduced to initial artefact-free scans during training and inference. 
    Using the raw \textit{k}-space data of the reference images, we undersample them with the acceleration factor of 4, impose rigid motion of a moderate amplitude, and introduce mild Gaussian noise. Note how the distortions differ from those in the Natural Images, on which the common IQMs were developed.
    We adjusted brightness for viewer’s convenience.}
    \label{fig:examples-distortions}
\end{figure*}


The key goal of this study is to evaluate popular selected IQMs on MRI data.
Previous works~\cite{Mason2020medicalcomparison, Renieblas2017medicalssim} evaluated the ability of certain IQMs to assess overall quality of data after to various types of artificial distortions\footnote{These artificial distortions, \textit{e.g.}, blurring or JPEG artifacts, are rarely encountered or even impossible in MRI practice.
}.
However, in practice, the overall image quality rating may be insufficient due to its ambiguity: \textit{e.g.}, one could not truly interpret the reasons for poor or good scoring. 
At the same time, asking the medical experts these general questions may be challenging because of many factors, ranging from the specifics of certain clinical workflows to personal preferences.

In this work, we aspire to solve these problems by proposing the following study. 
First, we evaluate IQMs with regard to their ability to reflect radiologists' perception of the quality of distorted images, comparing them to the fully-sampled artifact-free ones.
We range the metrics based on three IQ criteria that are crucial for making clinical decisions: perceived level of noise (SNR), perceived level of contrast (CNR), and the presence of artifacts.
Second, instead of corrupting images with artificial perturbations, for the first time in the community, we validate these metrics using the actual outputs of deep learning networks trained to solve common MR-related tasks.
As such, the artifacts originate from the imperfect solutions to the common real-world problems of motion correction, scan acceleration, and denoising.

A group of trained radiologists rated the quality of distorted images compared to the clean reference images on a scale from 1 to 4 for the three IQ criteria.
\myRed{Unlike the five-point Likert scale, the simplified scale balances the descriptiveness of the score with the noise in the votes of the radiologists. 
Our mock experiments showed that the respondents considered the selection between too many options difficult, with the five-point scale having a diluted difference between the options; whereas, the three-point scale was deemed insufficient.
}

\myRed{After the evaluation}, the aggregated results were compared with the values of selected IQA algorithms to identify the top performers -- the metrics that correlate the highest with the radiologists' votes.

\subsection{Image Library Generation}




As a data source, we use the largest publicly available repository of raw multi-coil MRI \textit{k}-space data -- the FastMRI dataset, containing the knee and the brain scans~\cite{KnollfastMRI2020challenge, zbontar2019fastmri_both_datasets}.
The knee subset of FastMRI contains 1,500 fully sampled MRIs acquired with a 2D protocol in the coronal direction with 15 channel knee coil array on 3 and 1.5~Tesla Siemens MRI machines.
The data consists of approximately equal number of scans acquired using the proton density weighting with (PDFS) and without (PD) fat suppression pulse sequences with the pixel size of 0.5~mm $\times$ 0.5~mm and the slice sickness of 3~mm.

The knee subset is divided into 4 categories: train (973 volumes), validation (199 volumes), test (118 volumes), and challenge (104 volumes). 
Only the multi-coil scans were selected for this study, omitting the single-coil data.

The brain subset includes 6,970 1.5 and 3 Tesla scans collected on Siemens machines using T1, T1 post-contrast, T2, and FLAIR acquisitions. 
Unlike the knee subset, this data are of a wide variety of reconstruction matrix sizes.
For the purpose of de-identification, authors of the dataset limited the data to only 2D axial images, and replaced \textit{k}-space slices $\gtrapprox$ 5 mm below the orbital rim with zero matrices.
The brain subset is divided into 6 categories: train (4,469 volumes), validation (1,378 volumes), test 4$\times$ (281 volumes), test 8$\times$ (277 volumes), challenge 4$\times$ (303 volumes), and challenge 8$\times$ (262 volumes).

Starting with the clean knee and brain data, we fist generate images corrupted with three types distortions: scan acceleration, motion, and noise.
The examples of the distorted images are presented in Fig. \ref{fig:examples-distortions}.
\myRed{After that, we train two reconstruction models for each type of distortions using PyTorch \cite{Paszke2019pytorch} for 2 and 4 epochs respectively.}
\myRed{The reduced training time was a conscious choice, enabling the model to produce some reconstruction errors. More specifically, we interrupted the training when the 90\% of the loss plateau is reached, which allows for good performing models with imperfections we wanted to test for\footnote{It is a standard way to broaden the image distribution from which the samples are drawn for evaluation and voting (\textit{e.g.}, see \cite{effland2019optimal}).}.}
\myRed{Finally, we use the trained models to reconstruct the corrupted images in the \textit{validation subset} of the fastMRI, from which we generate the labeling dataset.}

\medskip
\subsubsection{Scan Acceleration}

Scan-acceleration data are generated from the ground truth images by undersampling the \textit{k}-space data. 
To train the model, we selected only T1 weighted scans (T1, T1-PRE and T1-POST) from the train category of the FastMRI brain data.
The same subset of data was used for training of motion correction and denoising models.
The \textit{k}-space data were subsampled using a Cartesian mask, where \textit{k}-space lines are set to zero in the phase encoding direction.
The sampled lines are selected randomly, with the total sampling density depending on the chosen acceleration rate.
Following the data generation process from the FastMRI challenge \cite{KnollfastMRI2020challenge}, all masks are fully sampled in the central area of \textit{k}-space (the low frequencies). 
For the 4$\times$ accelerated scans, this corresponds to 8\%, and for the 8$\times$ acceleration, it equals to 4\%.
Besides making the reconstruction problem easier to solve, such lines allow computing the low-pass filtered versions of the images for assessing the coil sensitivity maps.

To compensate for the undersampling, we used the 2019 FastMRI challenge winner Adaptive-CS-Net model~\cite{pezzotti2020adaptivecsnet}. 
Based on the Iterative Shrinkage-Thresholding Algorithm (ISTA)~\cite{beck2009ista_framework}, this model consists of several trainable convolutional multi-scale transform blocks between which several prior knowledge-based computations are implemented.
For scalability reasons and without substantially impacting the reconstruction results, in this study, we trained a simplified light-weight version of the Adaptive-CS-Net model.
The resulting model consists of only 10 trainable blocks and 267k parameters.
\myRed{Unlike the full Adaptive-CS-Net model with three MRI-specific physics-inspired priors, the simplified version has only one prior module between the reconstruction blocks -- the soft data consistency step.}
\myRed{Specifically, the update for the block $B_{i+1}$ in the simplified Adaptive-CS-Net model is defined as follows:}

\begin{equation}
    B_{i+1}(\textbf{x}_i) = \textbf{x}_i + \hat{\mathcal{U}}_i(\textit{soft}(\mathcal{U}_i(\textbf{x}_i, \textbf{e}_{i}), \lambda_{s, f_s}))
    \label{eq:scan-acc-model-block}\,,
\end{equation}
\myRed{where $\textbf{x}_i$ denotes the $i$-th estimate of reconstruction, $\mathcal{U}$ and $\mathcal{\hat{U}}$ are the multi-scale transform and its inverse that consist of 2D convolutions and a nonlinearity in the form of Leaky-ReLU.
The feature maps produced at the different scales are thresholded
using the soft-max function $\textit{soft}(\cdot)$\footnote{Defined as $\textit{soft}(\textbf{u}, \lambda)=\max(|\textbf{u}| -\lambda, 0) \cdot \frac{\textbf{u}}{|\textbf{u}|}$.}, parameterized by a learned parameter $\lambda_{s, f_s}$ for each feature channel $f_s$ and scale $s$.
In Eq.~\ref{eq:scan-acc-model-block}, the soft data consistency step $ \textbf{e}_{i}$ is defined as follows:}

\begin{equation}
    \textbf{e}_{i} = \mathcal{F}^{-1}(M\mathcal{F}\textbf{x}_i - M\textbf{y}),
    \label{eq:soft-dc}
\end{equation}
\myRed{where $\mathcal{F}$ and $\mathcal{F}^{-1}$ denote Fourier transform and its inverse, $My$ is the data measured with the sampling mask $M$.}

\myRed{We trained the simplified Adaptive-CS-Net model using RMSprop optimizer~\cite{tieleman2012lecture} to minimize L1 loss function between the reconstruction estimate and the ground truth image obtained from the fully sampled data.
We used a step-wise learning rate decay of $10^{-4}$ \myRed{and the batch size of 8} to reconstruct the data for various acceleration factors (from 2$\times$ to 8$\times$).}

\medskip
\subsubsection{Motion Correction}

\begin{figure*}[!ht]
    \centering
    \includegraphics[width=\textwidth]{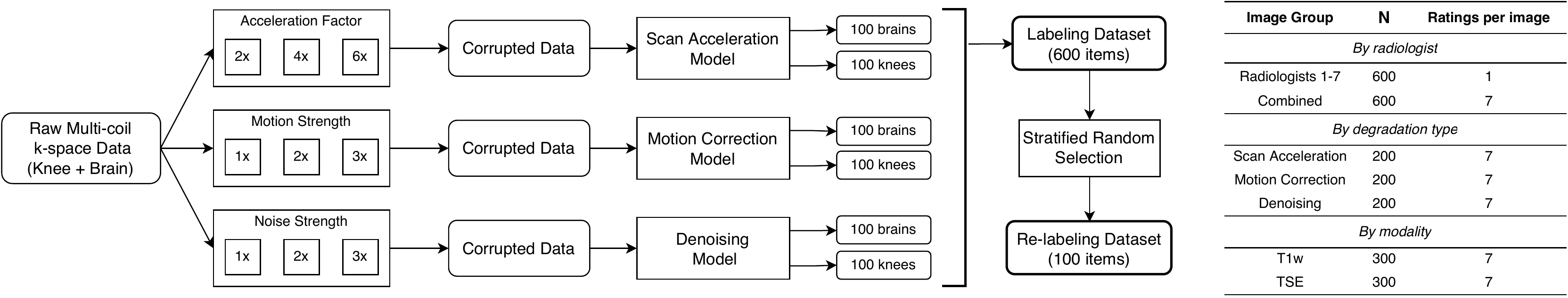}
    \caption{Formation of Labeling and Re-labeling datasets for annotation (left) and the content of each group of images for the medical evaluation and labeling by radiologists in the Labeling dataset (right). 
    Starting from the clean validation data, we first generate corrupted data with the acceleration artefacts, the motion artefacts, and the Gaussian noise. Then, we reconstruct the corrupted data using trained neural network models and randomly select scans to form labeling and re-labeling pairs for the experts to grade.}
    \label{fig:dataset-formation-and-table}
\end{figure*}

The in-plane motion artifacts, including rigid translation and rotation, were introduced into the Fourier-transformed data following the procedure described in \cite{sommer2020correction}.
For each input image, the assumed echo-train length of the turbo spin-echo readout was chosen randomly in the 8–-32 range. 
Similarly, the assumed extent of zero-padding in \textit{k}-space was chosen randomly in the range of 0–100.
The motion trajectories (translation/rotation vectors as a function of scan time) were generated randomly to simulate the realistic artifacts.
In this study we utilized the protocol for “sudden motion” simulation.
Here, the subject is assumed to lie still for a large part of the examination, until a swift translation or rotation of the head occurs. 
The time point of the sudden motion was taken randomly as a fraction of the total scan time in the range of one-third to seven-eighths.
\myRed{The maximum magnitude of the motion was chosen randomly from the range of [1,4] pixels for the translation and [0.5,4.0] degrees for the rotation artifacts. The center of rotation was also varied randomly in the range of [0,100] pixels in each direction.
These parameter ranges were selected empirically to generate a large variety of realistic artifacts and were used consistently in the training and in the validation runs.}

To compensate for the motion artefacts of various extent, we trained U-Net models~\cite{renneberger2015unet} \myRed{with 209k parameters}.
\myRed{While more advanced architectures exist, we found the basic U-Net to be more than sufficient for the scope of the proposed IQA study, as it is enough to capture imperfections which are often generated by deep learning models.}

The model received the motion corrupted data as the input and learned to predict the motion artefacts in a residual manner, \textit{i.e.}, the output of the model was a predicted image of motion in the input data.
The model was trained to minimise L1 loss between the ground-truth and the predicted residual with Adam~\cite{kingma2015adam} optimizer using the step-wise learning rate decay of $10^{-4}$ \myRed{and the batch size of 8}.
Preserving the same nature of artifacts, we trained our models for a range of amplification factors (from 1 to 3).
\myRed{For that, throughout the training, the motion amplitude was scaled by the amplification factor, yielding a consistently diverse appearance of the motion artefacts that could be met in practice.}

\medskip
\subsubsection{Denoising}


\myRed{In our study, noisy \textit{magnitude} images are generated from the complex \textit{k}-space data with the Gaussian distribution taken as the representative noise model.
Below, the standard deviation of the Gaussian noise is reported for a region of interest in the background of the magnitude image, as proposed in \cite{sijbers2007noise}}.
The parameters of the noise distribution for each volume are drawn from the last slice of this volume.
Then, the Gaussian noise with the estimated distribution parameters is \myRed{generated}, scaled by an amplification factor, and added to all images of the volume.
We used the amplification factor of 2 for the training and the amplification factors of 1, 2, and 3 for the test data generation \myRed{to enrich the variety of the tested image qualities in the resulting dataset}.

To compute the denoised images, we trained DnCNN models~\cite{dong2015image} \myRed{with 556k parameters} on the brain multi-coil train data using the RMSprop optimizer~\cite{tieleman2012lecture} and a step-wise learning rate decay of $10^{-4}$ \myRed{with the batch size of 8}.
\myRed{Similarly to the other tasks considered herein, we are not looking for the most powerful denoising algorithms but consider a very commonplace model DnCNN instead, merely to rank the modern IQA metrics for the specific task of denoising.}

\medskip
\subsubsection{Final Dataset for Labeling}

We started the formation of the labeling dataset from the clean volumes from the validation subsets of brain and knee FastMRI datasets; hence, these scans were not used to train the artefacts correction models.
In total, both validation subsets contain 1,577 volumes, resulting in 28,977 images: 199 knee volumes with 7,135 slices and 1,378 brain volumes with 21,842 slices.
In each brain volume, the lower 2 and top 3 slices were discarded to restrict the analysis to clinically relevant parts of the scan.
In each knee volume, the first 3 slices were discarded for the same reason.
To limit the number of data points and decrease the overall variability of data types, we selected only T1-weighted (T1, T1-PRE and T1-POST) brain volumes and proton-density weighted without fat suppression (PD) knee volumes.


The data generation pipeline is summarized in Fig.~\ref{fig:dataset-formation-and-table}.

Using the selected subset of clean validation data, we simulated images for the reconstruction:
\begin{itemize}
    \item For the scan acceleration task, we simulated acceleration artefacts for undersampling rates of 2$\times$, 4$\times$, and 6$\times$, following the data generation process from the FastMRI challenge;
    \item For the motion correction task, we simulated motion artefacts of three different strengths using the rigid motion simulation framework described above;
    \item For the denoising task, we simulated Gaussian noise with amplification factors of 1, 2 and 3 using the noise generation procedure described above.
\end{itemize}


After that, all generated corrupted data were reconstructed using the reconstruction models trained for the corresponding tasks. 
Note that we deliberately generated a fraction of data with parameters different from the ones used to train the reconstruction models. 
We found this approach yields various levels of artefacts typically appearing after the reconstruction process.

From the large pool of reconstructed images, we select 100 pairs of images (clean - reconstructed) for each task (scan acceleration, motion correction, denoising) and each anatomy (knee, brain), evenly distributing the data to represent each reconstruction parameter (\textit{e.g.}, the acceleration rate for the scan acceleration task).
This strategy results in the labeling dataset of 600 pairs of images in total (3 tasks $\times$ 2 anatomies).

To reach the goal labeling dataset size, we utilized the following data selection procedure:
\begin{enumerate}
    \item Compute values of IQMs for all reconstructed images (for NR IQMs) or image pairs (for FR IQMs);
    \item Normalize each IQM value to $[0, 1]$;
    \item Compute variance between IQM values for all items;
    \item Sort all items by the value of variance;
    \item Select 25\% of data for each task-anatomy combination from the data items with the highest variance, assuming that items with the biggest disagreement between IQMs are the most informative;
    \item Select the rest 75\% of data pseudo-randomly (preserve distribution of reconstruction parameters) to avoid introducing any bias from the variance computation.
\end{enumerate}

Lastly, we deliberately duplicated 100 of the 600 prepared items for the purpose of verification of radiologists' self-consistency, resulting in 700 image pairs to be labelled by each radiologist.

\subsection{Experiment Setup}

Within the paradigm of the model observer framework \cite{Barrett1993modelobservers}, the quality of a medical image can be defined as how well a clinical task (\textit{e.g.}, diagnostics) can be performed on it \cite{He2013modelobserversmedical}.
This means that the perfect MRI IQM would be some task-based score, such as the diagnostic accuracy.
However, such a metric is difficult to implement due to a great diversity of diagnostic outcomes that radiologists deal with in practice.
Because of that, the convention is to use a \textit{subjective estimation} of the overall diagnostic value instead \cite{Mason2020medicalcomparison}.

However, we argue that a single score is not sufficient to reflect the abundance of anatomies, pathologies, and artefactual cases that the radiologists work with.
Instead, we propose to subdivide the score of the overall diagnostic quality into three main criteria that can be important for a clinical practitioner to make their decision:
i) perceived level of noise, ii) perceived level of soft-tissue contrast, and iii) presence of artefacts.

\medskip
\subsubsection{Subjective Evaluation}
Seven trained radiologists \myRed{with 7 to 20 years of experience} took part in this study.
The participants were asked to score pairs of reconstructed-reference images using three main IQ criteria.
For each image pair and each criterion, radiologists scored the perceived diagnostic quality of the reconstructed image compared to the ground-truth using a four-point scale: not acceptable (1), weakly acceptable (2), rather acceptable (3), and fully acceptable (4).
The four-point scale was selected over the five-point Likert scale, previously used in \cite{Mason2020medicalcomparison}.

\begin{figure*}[!ht]
    \centering
    \includegraphics[width=0.9\linewidth]{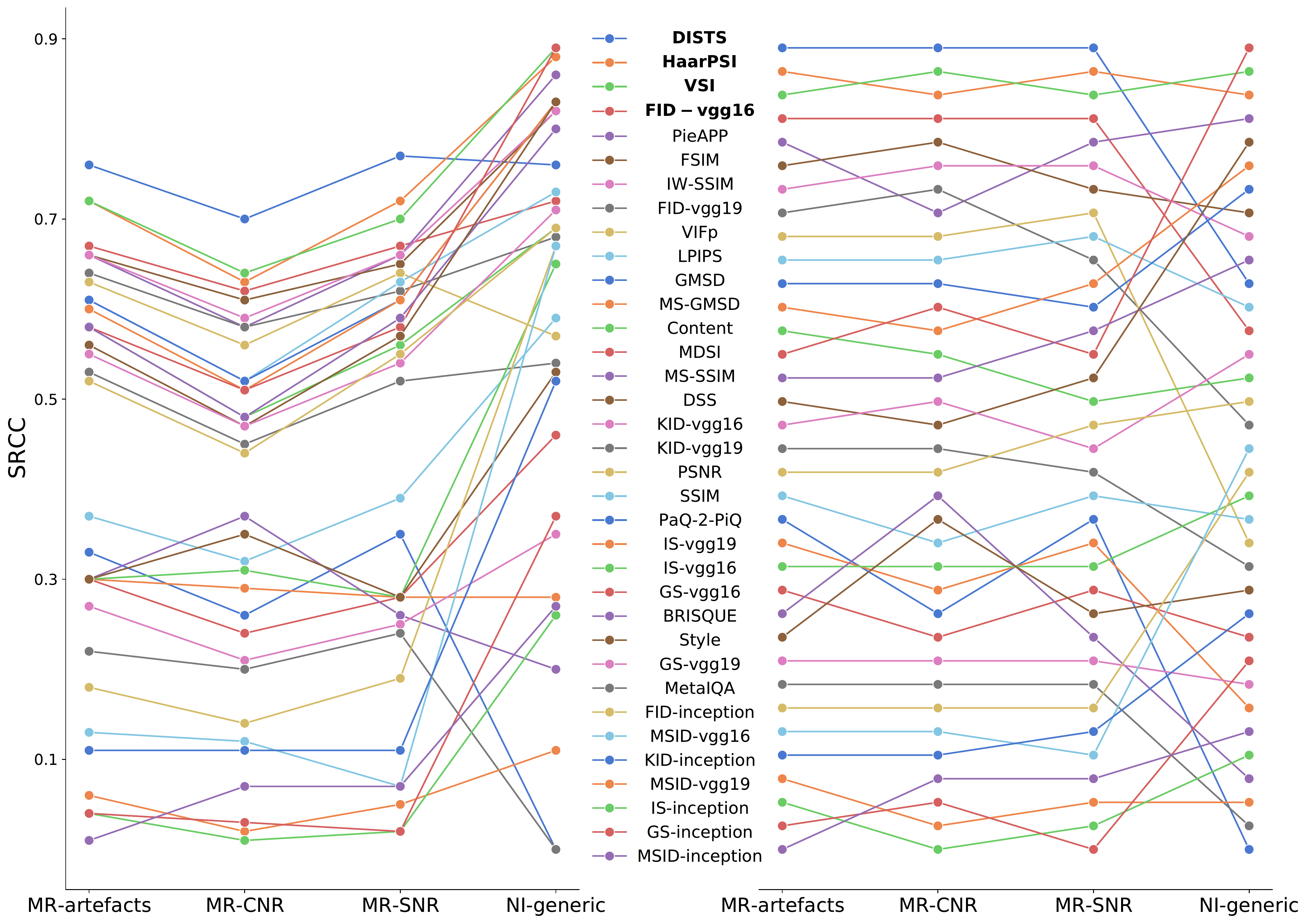}
    \caption{Performance of IQMs on different MRI tasks and on Natural Images (NI), compared by their correlation with the expert votes (SRCC values, left) and sorted top-to-bottom by their rank (right). 
    The ordering reflects the performance on the MRI data only. 
    The same color-coding is used in both plots. \textit{NI-generic} scores are the average between TID2013 \cite{ponomarenko2015image} and KADID-10k \cite{kadid10k} datasets. Note higher correlation of IQMs on NI and poor translation of ranking to MRI domain. 
    Refer to data in Table \ref{table:metrics_comparison_medical} for numerical values.
    }
    \label{fig:ni-mr-comparison}
\end{figure*}

Each participant performed the labelling individually using a dedicated instance of the Label Studio \cite{Label_Studio} software accessible via a web interface. The experts were asked to make all judgments about the image quality with regard to a particular diagnostic task that they would normally perform in their practice (\textit{e.g.}, the ability to discriminate relevant tissues, the confidence in using the image to detect a pathology, \textit{etc.}).
The interface provided additional functionality of scaling (zooming) the images to closer mimic the real-life workflow.
The pairs of images were displayed in a random order until all pairs were labelled.
Participants had an opportunity to re-label the pairs they have already scored at any point until the experiment is finished.

During the main part of the experiment, each participant labelled 600 pairs of images based on the 3 quality criteria, resulting in 4,200 annotated pairs and 12,600 labels in total.
The results of the main labelling session were used for further evaluation of the IQMs.
After finishing the main part of the experiment, the participants were asked to additionally label 100 randomly selected pairs from the same dataset, yielding additional 2,100 labels.
The results of this additional re-labelling were used to evaluate the self-consistency of each annotator.

\medskip
\subsubsection{Metrics Computation}

Unlike FR and NR IQMs, designed to compute an image-wise distance, the DB metrics compare distributions of \textit{sets} of images.
This makes them less practical for traditional IQA, the goal of which is to compute a score for a given image pair. 
Moreover, the need to have sets of images hinders the vote-based evaluation via the mean subjective opinion scores. 

To address these problems, we adopt a different way of computing the DB IQMs. 
Instead of extracting features from the whole images, we crop them into overlapping tiles of size $96 \times 96$ with $stride = 32$.
This pre-processing allows us to treat each pair of images as a pair of distributions of tiles, enabling further comparison. 
The other stages of computing the DB IQMs are kept intact.

\subsection{Data Analysis}


Here, we adapt the analysis of the scoring data proposed in \cite{Sheikh2006} to the multiple IQ criteria. 
The voting scores for each scoring criteria are not analyzed in their raw format. 
Instead, they are converted to z-scores (averaged and re-scaled from 0 to 100 for each radiologist to account for their different scoring):

\begin{equation}
    z_{nmk} = (D_{nmk} - \mu_{mk})/\sigma_{mk}\,,
    \label{eq:zscore}
\end{equation}

\noindent where $\mu_{mk}$ and $\sigma_{mk}$ are the mean and the standard deviation of the difference scores of the $m^{\text{th}}$ radiologist on the $k^{\text{th}}$ scoring criteria, and $D_{nmk}$ are the difference scores for $n^{\text{th}}$ degraded image defined as follows:

\begin{equation}
    D_{nmk} = s_{mk,\text{ref}} - s_{nmk}\,.
    \label{eq:difff}
\end{equation}
\noindent
In Eq.~(\ref{eq:difff}), $s_{mk,\text{ref}}$ is the raw score of the $m^{\text{th}}$ radiologist on the $k^{\text{th}}$ scoring criteria for the reference image corresponding to the $n^{\text{th}}$ degraded image, and $s_{nmk}$ is the raw score of the $m^{\text{th}}$ radiologist on the $n^{\text{th}}$ degraded image on the $k^{\text{th}}$ scoring criteria.
Note that in this study, the radiologists were asked to perform pair-wise comparison between degraded and reference images. 
Hence, it is possible to treat the raw labelling scores as the difference scores $D_{nmk}$.

After standardizing the expert votes by Eq.~(\ref{eq:zscore}), their correlation statistics with each IQM were computed in the form of SRCC and KRCC coefficients, \myRed{defined as follows:}

\begin{equation}
\text{SRCC} = 1 -  \frac{6 \sum_{i=1}^{n} d_i ^ 2}{n (n^2 - 1)},
\end{equation}
\noindent
\myRed{where $d_i$ is the difference between the \textit{i}-th image’s ranks in the objective and the subjective ratings and $n$ is the number of observations.}

\begin{equation}
\text{KRCC} = \frac{2}{n(n-1)}\sum_{i<j} \text{sign}(x_i-x_j)\;\text{sign}(y_i-y_j)\,,
\end{equation}
\noindent
\myRed{where $(x_1, y_1),...,(x_n, y_n)$ are the observations: the objective and the subjective score pairs.}

We use SRCC as the main measure of an IQM performance, due to the non-linear relationship between the subjective and the objective scores\footnote{The non-linear relationship is evident in Fig. \ref{fig:subjective-objective-relation} below.}.

The sizes of each batch of data are described in Fig. \ref{fig:dataset-formation-and-table} (right).

A non-linear regression was performed on the IQM scores according to the quality $Q$ to fit the subjective votes:

\begin{equation}
    Q(x) = \beta_1 \left( \frac{1}{2} - \frac{1}{1 + \exp(\beta_2 (x - \beta_3))} \right) + \beta_4 x + \beta_5,
    \label{eq:non-linear-regression}
\end{equation}

\noindent
where $x$ are the original IQM scores and $\beta_1$, $\dots$, $\beta_5$ are the fitting coefficients.
\section{Results}

\begin{figure*}[!ht]
    \centering
    \includegraphics[width=\textwidth]{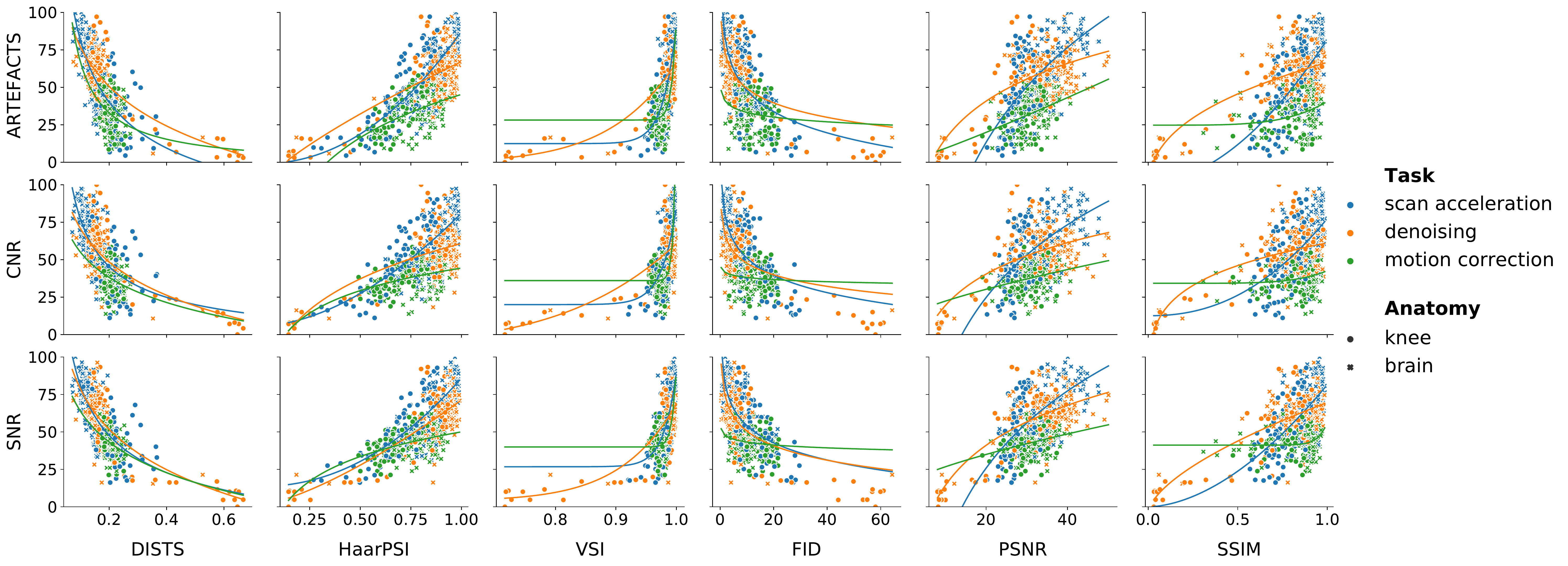}
    \caption{Relationship between processed subjective scores and IQM values for 3 evaluation criteria, 3 target tasks, and 2 anatomies (600 annotated image pairs in total).
    The solid lines are fits, plotted using the non-linear regression (\ref{eq:non-linear-regression}) on the subsets of images split by the tasks.
    The top 4 metrics (along with PSNR and SSIM, as the most commonplace) are shown in the decreasing order left to right, using SRCC to gauge the performance.
    }
    \label{fig:subjective-objective-relation}
\end{figure*}

\begin{figure*}[!ht]
    \centering
    \includegraphics[width=0.95\linewidth]{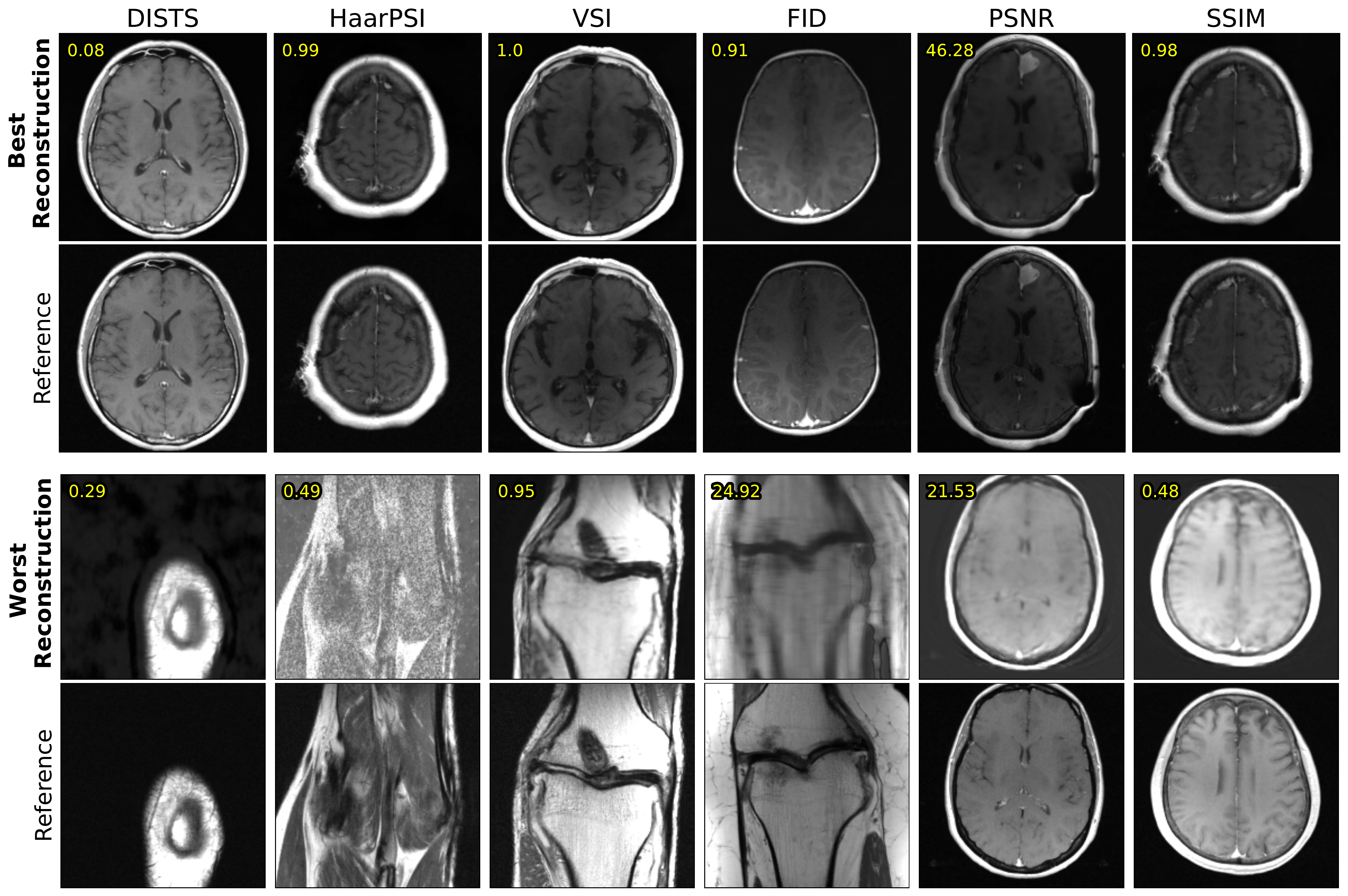}
    \caption{The best and the worst reconstruction-reference pairs according to different metrics
    (their values are shown in yellow). 
   Note how the top 4 metrics (first four columns) reflect the actual reconstruction quality better than PSNR and SSIM (which are prone to \textbf{misjudging a simple shift of brightness or a blur}).
    The brightness is adjusted for viewer's convenience.}
    \label{fig:best-worst-recons-by-different-metrics}
\end{figure*}

Figs. \ref{fig:ni-mr-comparison} and \ref{fig:subjective-objective-relation} and Table \ref{table:metrics_comparison_medical}
summarize the correlation study between the radiologists’ scores and the IQM values for the three proposed evaluation criteria.
\myRed{The figures also show the results for the natural image domain.}
Top 4 performers in each category are marked in bold. 
The best and the worst examples of the reconstructions, as judged by different metrics, are presented in Fig. \ref{fig:best-worst-recons-by-different-metrics}, and the aggregate scores for the top-performing metrics in each application in Fig. \ref{fig:subjective-objective-relation-combined}.

%

\section{Discussion}

The visual inspection of the outputs of the models in Fig. \ref{fig:best-worst-recons-by-different-metrics} makes it evident how the top metrics are superior in reflecting the actual reconstruction quality over the conventional PSNR and SSIM. The latter are known to misjudge shifts of brightness or a blur, indicating high quality for the bad images, whereas the more advanced FR and DB IQMs correlate with the visual perception and the subjective scores ably.  
Henceforth, out of the 35 metrics considered, we only discuss the best ones, according to their rank in the correlation study (VSI, HaarPSI, DISTS, FID\textsubscript{VGG16}) 
and the widely used PSNR and SSIM.

As the key observation in the first systematic study of the DB metrics, we affirm that the choice of the feature extractor plays a crucial role.
In particular, the correlation scores show that the Inception-based features are almost always worse than those from VGG16 (except for the MSID metric).
Moreover, we see that, despite having been designed for the evaluation of \textit{realism} of generative models data, FID shows competitive SRCC scores, thus, becoming a new recommended metric for the MRI image assessment tasks.

\begin{table}[tbh!]
\caption{SRCC values of all 35 metrics on Natural and MRI Data. \\ Top 4 performers in all categories are marked in bold. \\$^*$ denotes values taken directly from \cite{Golestaneh2021noreferencereview}}
\vspace{3mm}
\setlength\tabcolsep{2.5pt}
\resizebox{1\columnwidth}{!}{ 
\begin{tabular}{lcccccc}
\toprule

&\multicolumn{2}{c}{Natural Images}&&\multicolumn{3}{c}{MRI Data}  \\
\cmidrule{2-3} \cmidrule{5-7}
& TID2013 & KADID-10k && Artefacts & CNR & SNR \\

  \cmidrule{2-7}
  PSNR & $0.69$ & $0.68$ && $0.52$ & $0.44$ &  $0.55$       \\
  SSIM \cite{wang2004image} & $0.55$ & $0.63$ && $0.37$ & $0.32$ & $0.39$  \\
  MS-SSIM \cite{wang2003multiscale} & $0.80$ & $0.80$ && $0.58$ & $0.48$ & $0.59$  \\
  IW-SSIM \cite{wang2010information} & $0.78$ & $0.85$ && $0.66$ & $0.59$ & $0.66$ \\
  VIFp \cite{sheikh2005visual} & $0.46$ & $0.67$ && $0.63$ & $0.56$ & $0.64$  \\
  GMSD \cite{xue2013gradient} & $0.80$ & $0.85$ && $0.61$ & $0.52$ & $0.61$ \\
  MS-GMSD \cite{Zhang2017gmsd} & $0.81$ & $0.85$ && $0.60$ & $0.51$ & $0.61$  \\
  FSIM \cite{zhang2011fsim} & $0.80$ & $0.84$ && $0.66$ & $0.61$ & $0.65$  \\
  VSI \cite{zhang2014vsi} & $\mathbf{0.89}$ & $\mathbf{0.88}$ && $\mathbf{0.72}$ & $\mathbf{0.64}$ & $\mathbf{0.70}$ \\
  MDSI \cite{nafchi2016mdsi} & $\mathbf{0.89}$ & $\mathbf{0.89}$ && $0.58$ & $0.51$ & $0.58$ \\
  HaarPSI \cite{Reisenhofer2018haarpsi} & $\mathbf{0.87}$ & $\mathbf{0.88}$ && $\mathbf{0.72}$ & $\mathbf{0.63}$ & $\mathbf{0.72}$ \\
  Content\textsubscript{VGG16} \cite{johnson2016perceptual} & $0.67$ & $0.71$ && $0.58$ & $0.48$ & $0.56$  \\
  Style\textsubscript{VGG16} \cite{johnson2016perceptual} & $0.50$ & $0.56$ && $0.30$  & $0.35$ & $0.28$ \\
  LPIPS\textsubscript{VGG16} \cite{zhang2018unreasonable} & $0.67$ & $0.78$ && $0.61$ & $0.52$ & $0.63$  \\
  DISTS \cite{ding2020image} & $0.71$ & $0.81$ && $\mathbf{0.76}$ & $\mathbf{0.70}$ & $\mathbf{0.77}$ \\
  PieAPP \cite{Prashnani2018pieapp} & $0.84$ & $\mathbf{0.87}$ && $0.66$ & $0.58$ & $0.66$ \\
  DSS \cite{Balanov2015dss} & $0.79$ & $0.86$ && $0.56$ & $0.47$ & $0.57$ \\

\cmidrule{1-7}
\multicolumn{5}{c}{No-reference metrics} \\
\cmidrule{1-7}
  BRISQUE \cite{MittalMB12} & $0.20$ & $0.20$ && $0.30$ & $0.37$ & $0.26$  \\
  PaQ-2-PiQ \cite{paq2piq2020} & $\mathbf{0.86^*}$ & $0.84^*$ && $0.33$ & $0.26$ & $0.35$  \\
  MetaIQA \cite{MetaIQA2020} & $\mathbf{0.86^*}$ & $0.76^*$ && $0.22$ & $0.20$ & $0.24$ \\

\cmidrule{1-7}
\multicolumn{5}{c}{Distribution-based metrics} \\
\cmidrule{1-7}
  KID\textsubscript{InceptionV3} \cite{binkowski2018demystifying} & $0.42$ & $0.63$ && $0.11$ & $0.11$ & $0.11$  \\
  FID\textsubscript{InceptionV3} \cite{heusel2017gans} & $0.67$ & $0.66$ && $0.18$ & $0.14$ & $0.19$  \\
  GS\textsubscript{InceptionV3} \cite{khrulkov2018geometry} & $0.37$ & $0.37$ && $0.04$ & $0.03$ & $0.02$ \\
  IS\textsubscript{InceptionV3} \cite{salimans2016improved} & $0.26$ & $0.25$ && $0.04$ & $0.01$ & $0.02$ \\
  MSID\textsubscript{InceptionV3} \cite{tsitsulin2020shape} & $0.21$ & $0.32$ && $0.01$ & $0.07$ & $0.07$ \\
  
  KID\textsubscript{VGG16} \cite{binkowski2018demystifying} & $0.70$ & $0.71$ && $0.55$ & $0.47$ & $0.54$ \\
  FID\textsubscript{VGG16} \cite{heusel2017gans} & $0.67$ & $0.66$ && $\mathbf{0.67}$ & $\mathbf{0.62}$ &  $\mathbf{0.67}$ \\
  GS\textsubscript{VGG16} \cite{khrulkov2018geometry} & $0.47$ & $0.45$ && $0.30$ & $0.24$ &  $0.28$ \\
  IS\textsubscript{VGG16} \cite{salimans2016improved} & $0.64$ & $0.65$ && $0.30$ & $0.31$ &  $0.28$ \\
  MSID\textsubscript{VGG16} \cite{tsitsulin2020shape} & $0.69$ & $0.64$ && $0.13$ & $0.12$ & $0.07$ \\
  
  KID\textsubscript{VGG19} \cite{binkowski2018demystifying} & $0.54$ & $0.59$ && $0.53$ & $0.45$ & $0.52$ \\
  FID\textsubscript{VGG19} \cite{heusel2017gans} & $0.68$ & $0.75$ && $0.64$ & $0.58$ &  $0.62$ \\
  GS\textsubscript{VGG19} \cite{khrulkov2018geometry} & $0.35$ & $0.41$ && $0.27$ & $0.21$ & $0.25$ \\
  IS\textsubscript{VGG19} \cite{salimans2016improved} & $0.28$ & $0.33$ && $0.30$ & $0.29$ & $0.28$ \\
  MSID\textsubscript{VGG19} \cite{tsitsulin2020shape} & $0.11$ & $0.13$ && $0.06$ & $0.02$ & $0.05$ \\

\bottomrule
\end{tabular}
} 
\label{table:metrics_comparison_medical}
\end{table}

The non-linear relationship between the subjective and the objective scores, seen in Fig.~\ref{fig:subjective-objective-relation}, portrays intricate behavior with evident dependence on the anatomy and the target task, as well as a clear clustering of the points, instrumental for selecting a proper metric in a particular application.
Notable are the generally lower IQM correlation scores when the 
difficulty of the reconstruction routine increases (compare trends in the scan acceleration data to those in the more complex denoising and the motion correction models).
Also, the evaluation values for the knee reconstruction are generically lower, which could be caused by the greater variety of anatomical structures present in the knee data, as well as the more strict pertinent medical evaluation criteria \cite{Keshavan2019expertassessment}.

\definecolor{myblue}{RGB}{32,119,180}
\newcommand{\bluecircle}{\raisebox{0.5pt}{\tikz{\node[draw,scale=0.4,circle,fill=myblue](){};}}}
\newcommand{\greensquare}{\raisebox{0.5pt}{\tikz{\node[draw,scale=0.4,regular polygon, regular polygon sides=4,fill=black!35!green](){};}}}
\definecolor{myorange}{RGB}{255,127,15}
\newcommand{\orangetriangle}{\raisebox{0pt}{\tikz{\node[draw,scale=0.3,regular polygon, regular polygon sides=3,fill=myorange,rotate=180](){};}}}

\begin{figure}[h]
    \centering
    \includegraphics[width=\linewidth]{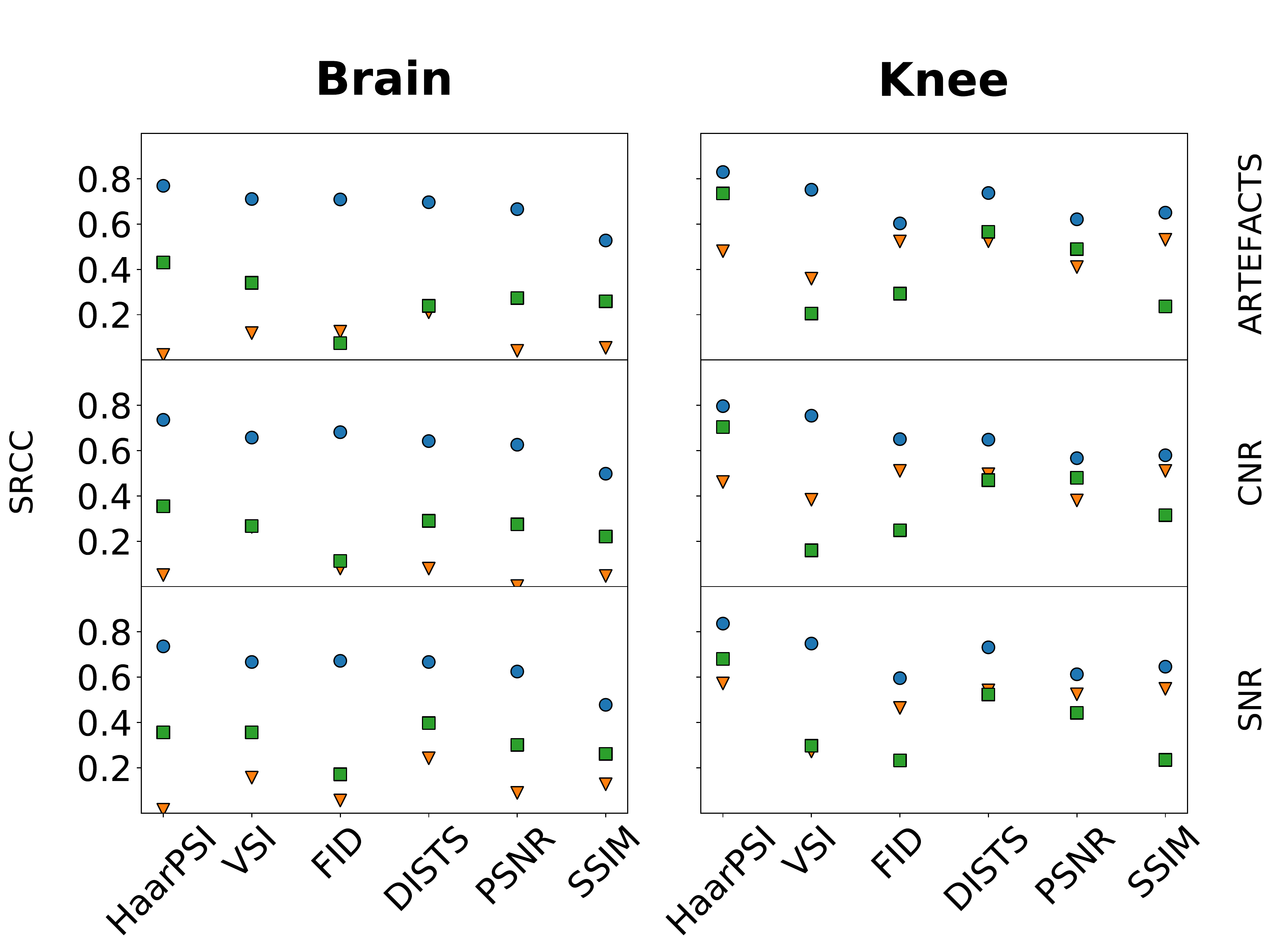}
    \caption{Aggregate relationship between the objective and the subjective scores for 3 evaluation criteria (rows), 2 anatomies (columns), and 3 tasks: scan acceleration (\,\protect\bluecircle\,), denoising (\,\protect\orangetriangle\,), and motion correction (\,\protect\greensquare\,). 
    The IQMs are ordered by decreasing average SRCC for the artefacts criterion on the brain data. 
    This order is kept throughout all results for consistency. 
    Note the tendency of the metrics to perform poorly in some task-anatomy combinations, \textit{e.g.}, in denoising the brain data.}
    \label{fig:subjective-objective-relation-combined}
\end{figure}

Fig. \ref{fig:subjective-objective-relation-combined} aggregates the outcomes per each task, anatomy, and evaluation criteria studied in our work, with the relation between the subjective and the objective scores highlighting the differences in the average performance of the top metrics. 
Notably, these selected IQMs have the highest correlation with expert judgment in the scan acceleration task. However, all metrics equally struggle reflecting the opinion of the radiologists in denoising and, sometimes, in motion correction tasks, especially on brain data.
We also observe that some metrics perform consistently in terms of all three evaluation criteria and all tasks for given anatomy.
For instance, GMSD and DISTS, despite not being of the highest SRCC rank overall, still show consistently high correlation scores on knee data, which proffers both of them as universal choices for the IQA in orthopedic applications. On the other hand, HaarPSI consistently rates the highest for both anatomies in the scan acceleration task, an instrumental fact to know when a single machine is used to scan various body parts or when the pertinent cross-anatomy inference \cite{belovUltrafast} is performed.

\subsection{Natural \textit{vs.} MRI Images}

A frequent IQA-related question is how generalizable are the performance benchmarks across different datasets and image domains.
To study that, we analyzed the applicability of all 35 IQMs considered herein both in the MRI and the natural image (NI) domains (Table \ref{table:metrics_comparison_medical}).
For the latter, the popular TID2013 \cite{ponomarenko2015image} and KADID-10k \cite{kadid10k} datasets of NIs were used.
Fig. \ref{fig:ni-mr-comparison} illustrates the effect of the shift between the NI and the MRI domains, featuring an expected drop of the correlation values for most metrics\footnote{Not surprising, given these IQMs were designed for NI in the first place.}.
However, the domain shift affects the ranks of the IQMs differently. 
Some top NI metrics, such as MDSI and MS-GMSD, naturally take lower standings in the MRI domain; however, others, such as HaarPSI and VSI, remain well-correlated with the radiologists' perception of quality.
Further examples of IQMs robust to the domain shift are DISTS and FID\textsubscript{VGG16}.
%


\subsection{Labeling Discrepancies and Self-Consistency Study}
Another IQA-related question encountered in survey-based studies is the trustworthiness of the votes themselves.
Given that only reputable radiologists were engaged in our labeling routine, we have no grounds for doubting their annotations as far as the domain knowledge is concerned. 
Therefore, feasible discrepancies among their votes can be assumed to originate either from such factors as the study design, its duration, and fatigue, or from a previous experience which sometimes forms \textit{a posteriori} intuition and, allegedly, influences the experts to make decisions different from the others. 

\begin{figure}[t]
    \centering
    \includegraphics[width=\linewidth]{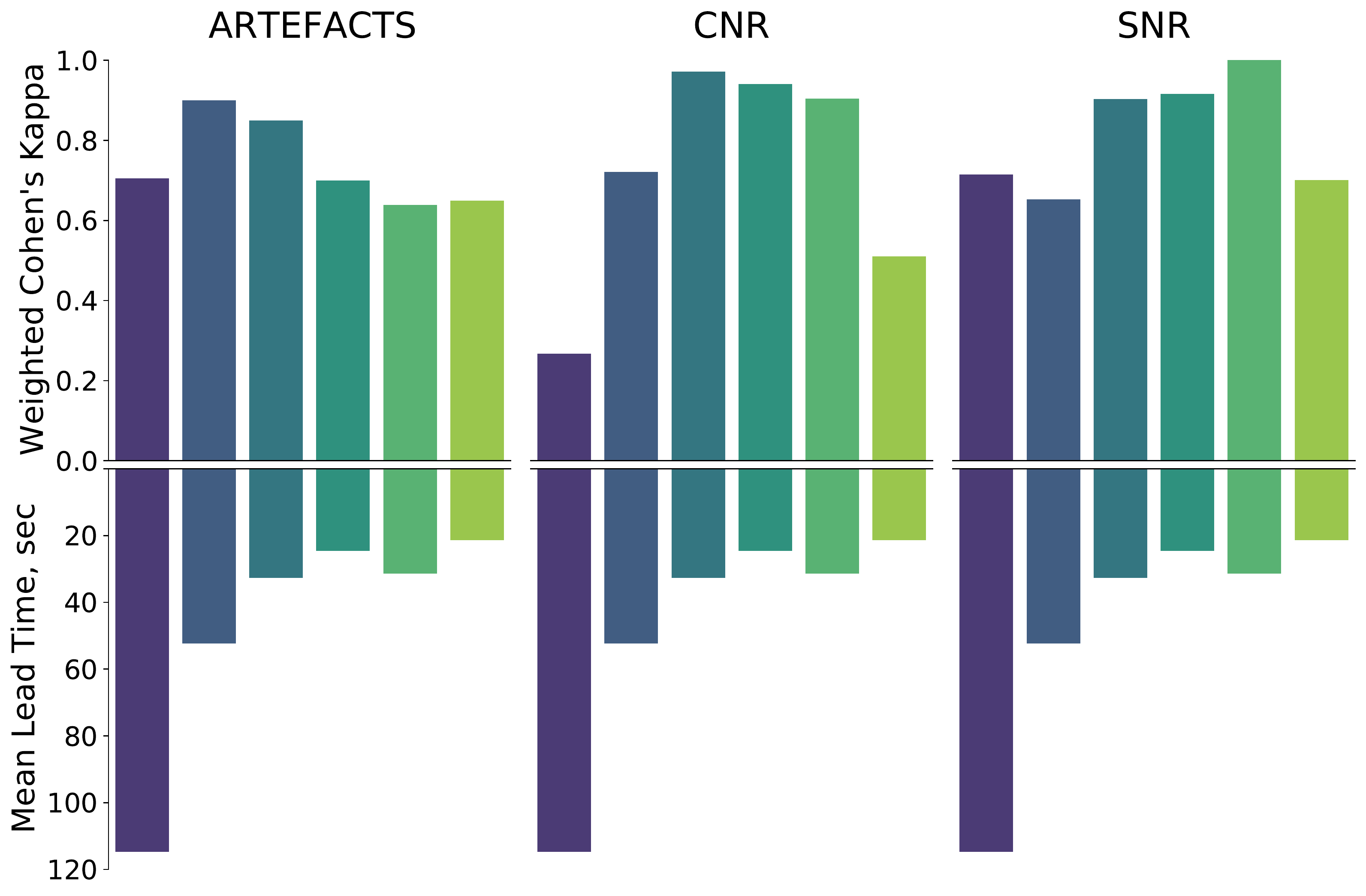}
    \caption{Correlation between the subjective scores in labelling and re-labelling sessions on the same data,
    with each column/color corresponding to an individual radiologist. 
    This plot shows scoring self-consistency of the experts and the average time spent labelling one pair of images.
    Apparently, the time spent on labelling is not the major factor affecting the self-consistency of experienced radiologists.
    }
    \label{fig:self-consistency}
\end{figure}

While the latter is too subjective and difficult to regulate, the former could be controlled. 
We put effort to simplify the user experience and allowed the radiologists to approach the labeling assignment in batches at their own pace.
The average lead time spent labeling a pair of images\footnote{We discarded 5\% of the shortest and the longest lead times to account for erroneous clicks and breaks between the labeling sessions.}, an arguable indicator of the scrupulousness of an annotator, is plotted in Fig.~\ref{fig:self-consistency}, where we also summarize the results of the self-consistency study.
The study reports Weighted Cohen's Kappa scores, computed between the votes provided in the main and in the additional re-labeling experiments on the same data.
Interestingly, there is no significant correlation between self-consistency and the labeling time, placing other factors mentioned above, such as individual experience, at the forefront.

In Fig.~\ref{fig:self-consistency}, the Weighted Cohen's Kappa values correspond to \textit{moderate to substantial} consistency of scoring (according to \cite{viera2005UnderstadingKappa}).
Given the sufficiently trustworthy labeling, the spread of the correlation scores for the modern IQA metrics in Fig. \ref{fig:subjective-objective-relation-combined}, and the non-trivial correlation patterns in Fig. \ref{fig:subjective-objective-relation}, one can conclude that \textit{the optimal} MRI metric is yet to be devised.

\smallskip
Besides a blunt umbrella metric aggregating the top-performing predictions (\textit{e.g.}, those of VSI, HaarPSI, DISTS, and FID\textsubscript{VGG16}), the future effort should be dedicated to additional forays into modeling \textit{MRI-specific perception} of the radiologists and to \textit{interpreting} their assessment using formalized rules taken from the medical textbooks. 
Such interpretatable metrics will be especially in demand, given the recent appearance of the MRI sampling approaches aimed towards optimizing downstream tasks \cite{razumov2021optimal}, including the recently annotated FastMRI dataset~\cite{fastmriplus}.
Another line of future work could be `borrowed' from the NI domain, where the abundance of data has led to the emergence of several NR IQMs. Although, in our study, all such metrics (classic BRISQUE \cite{MittalMB12} and the more recent PaQ-2-PiQ \cite{paq2piq2020} and MetaIQA \cite{MetaIQA2020}) showed equally mediocre performance compared to the other IQMs,
we believe their value in the MRI domain is bound to improve with the growth of available data.

\section{Conclusion}
\label{sec:conclusions}
This manuscript reports the most extensive study of the image quality metrics for Magnetic Resonance Imaging to date, evaluating 35 modern metrics and using 14,700 subjective votes from experienced radiologists.
\myRed{The top performers -- DISTS, HaarPSI, VSI, and FID\textsubscript{VGG16} -- are found to be efficient across three proposed quality criteria, for all considered anatomies and the target tasks.}
\section*{Acknowledgments}

We acknowledge the effort of radiologists from the Philips Clinical Application Team (PD CEER) for their help with data labeling and thank the supporters of our GitHub project\footnote{\href{https://github.com/photosynthesis-team/piq/}{https://github.com/photosynthesis-team/piq/}}, where each metric was independently implemented and tested. 
%

\bibliographystyle{model2-names.bst}\biboptions{authoryear}
\bibliography{main}

\newpage\section*{Appendix. Labelling User Interface}

\begin{figure}[h!]
    \centering
    \includegraphics[scale=0.3]{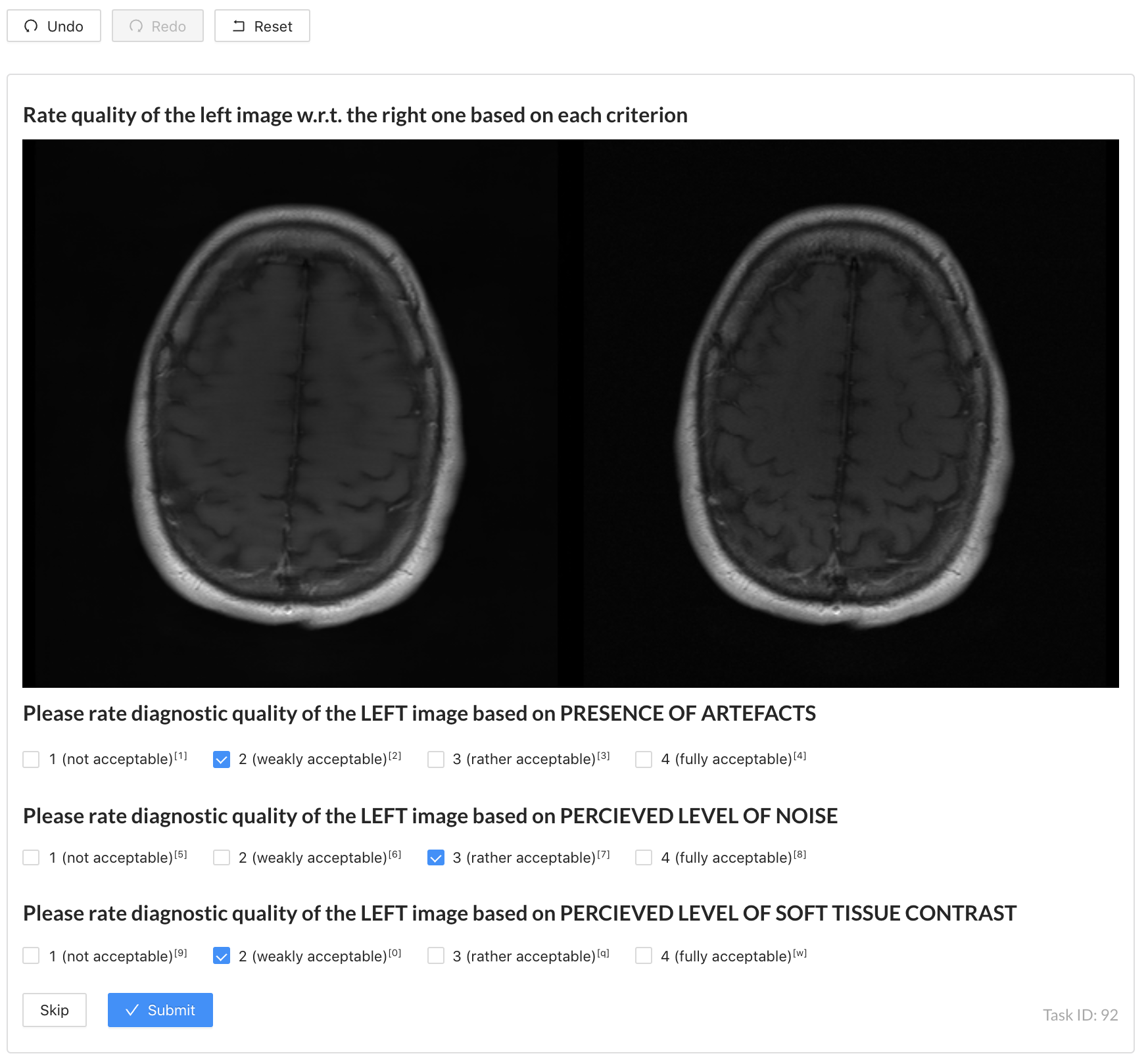}
    \caption{Web interface of the Label Studio software released to the expert radiologists to perform the labeling. The participants selected their answers using the proposed scale from 1 to 4, rating the images based on each proposed IQA criteria.}
    \label{fig:label-studio-interface}
\end{figure}

During the labeling experiment, the participants were asked to score pairs of reconstructed-reference images presented to them side-by-side in a web interface of the Label Studio \cite{Label_Studio}.
The web interface is shown in Fig. \ref{fig:label-studio-interface}.

The labeling was done using three main IQ criteria: the presence of artifacts, the perceived level of noise, and the perceived level of soft-tissue contrast.
The participants were able to select their answers using the mouse pointer or some keys on the keyboard.
During the quality assessment process, the participants were able to zoom images, re-label previously labeled examples, pause and divide their evaluation session into as many labeling rounds as they wished. 
All labeling results were continuously saved on a remote server to eliminate the possibility of data loss.
After the labeling process is finished, the participants were offered the last chance to fix the scoring of the borderline examples. 

\label{sec:labelling}

\end{document}